\documentclass{aa}  
\def \eg {e.g.}

\def \ie {i.e.}
\def \cf {cf.}
\def \lcdm {{\hbox{$\Lambda$CDM}}}
\def \omegam {{\hbox{$\Omega_m$}}}
\def \omegal {{\hbox{$\Omega_\Lambda$}}}
\def \hzero {{\hbox{$H_0$}}}
\def \arcmin {\hbox{$^\prime$}}
\def \arcsec {\hbox{$^{\prime\prime}$}}
\def \deg {\hbox{$^\circ$}}

\def \msun {\hbox{${\rm M_\odot}$}}

\def \mfive {\hbox{$M_{500}$}}
\def \lfive {\hbox{$L_{500}$}}
\def \rfive {\hbox{$r_{500}$}}

\newcommand{\ergs }{\mbox{erg s$^{-1}$}}

\newcommand{\kmsmpc }{\mbox{km s$^{-1}$ Mpc$^{-1}$}}

\newcommand{\kev }{\mbox{keV}}

\newcommand{\jy }{\mbox{Jy}}
\newcommand{\mjyb }{\mbox{mJy beam$^{-1}$}}
\newcommand{\mujyb }{\mbox{$\mu$Jy beam$^{-1}$}}

\newcommand{\whz }{\mbox{W Hz$^{-1}$}}

\newcommand{\uv }{\textit{uv}}

\newcommand{\prefactor }{\textsc{prefactor}}

\newcommand{\wsclean }{\textsc{WSClean}}

\newcommand{\killms }{\textsc{killMS}}
\newcommand{\ddfacet }{\textsc{DDFacet}}

\newcommand{\esas }{\textsc{esas}}
\newcommand{\esasE }{Extended Source Analysis Software}

\newcommand{\xmm }{{\em XMM-Newton}}

\newcommand{\gmrt }{GMRT}

\newcommand{\jvla }{JVLA}

\newcommand{\lofar }{LOFAR}
\newcommand{\lofarE }{LOw Frequency ARray}

\newcommand{\skaE }{Square Kilometer Array}

\newcommand{\lotss }{LoTSS}
\newcommand{\lotssE }{LOFAR Two-meter Sky Survey}

\newcommand{\tgss }{TGSS}
\newcommand{\tgssE }{TIFR GMRT Sky Survey}

\newcommand{\nvss }{NVSS}
\newcommand{\nvssE }{NRAO VLA Sky Survey}

\newcommand{\xcop }{X-COP}
\newcommand{\xcopE }{{\em XMM-Newton} Cluster Outskirts Project}
\usepackage{graphicx}
\usepackage{amssymb}
\usepackage{multirow,bigdelim}
\usepackage{txfonts}
\usepackage[export]{adjustbox}
\begin{document} 

\title{Particle acceleration in a nearby galaxy cluster pair: the role of cluster dynamics}

\authorrunning{A. Botteon et al.} 
\titlerunning{Particle acceleration in a nearby galaxy cluster pair: the role of cluster dynamics}

\author{A. Botteon\inst{\ref{unibo},\ref{ira},\ref{leiden}}, R. Cassano\inst{\ref{ira}}, D. Eckert\inst{\ref{geneva}}, G. Brunetti\inst{\ref{ira}}, D. Dallacasa\inst{1,\ref{ira}}, T. W. Shimwell\inst{\ref{leiden},\ref{astron}}, R. J. van Weeren\inst{\ref{leiden}}, F. Gastaldello\inst{\ref{iasf}}, A. Bonafede\inst{\ref{unibo},\ref{ira},\ref{hamburg}}, M. Br\"{u}ggen\inst{\ref{hamburg}}, L. B\^{\i}rzan\inst{\ref{hamburg}}, S. Clavico\inst{\ref{brera}}, V. Cuciti\inst{\ref{hamburg}}, F. de Gasperin\inst{\ref{hamburg}}, S. De Grandi\inst{\ref{brera}}, S. Ettori\inst{\ref{oas},\ref{infn}}, S. Ghizzardi\inst{\ref{iasf}}, M. Rossetti\inst{\ref{iasf}}, H. J. A. R\"{o}ttgering\inst{\ref{leiden}} and M. Sereno\inst{\ref{oas}}}

\institute{
Dipartimento di Fisica e Astronomia, Universit\`{a} di Bologna, via P.~Gobetti 93/2, I-40129 Bologna, Italy \\
\email{botteon@ira.inaf.it} \label{unibo}
\and
INAF - IRA, via P.~Gobetti 101, I-40129 Bologna, Italy \label{ira}
\and
Leiden Observatory, Leiden University, PO Box 9513, NL-2300 RA Leiden, The Netherlands \label{leiden}
\and
Department of Astronomy, University of Geneva, ch. d'Ecogia 16, 1290 Versoix, Switzerland \label{geneva}
\and
ASTRON, the Netherlands Institute for Radio Astronomy, Postbus 2, NL-7990 AA Dwingeloo, The Netherlands \label{astron}
\and
INAF - IASF Milano, via A.~Corti 12, I-20133 Milano, Italy \label{iasf}
\and
Hamburger Sternwarte, Universit\"{a}t Hamburg, Gojenbergsweg 112, D-21029 Hamburg, Germany \label{hamburg}
\and
INAF - Osservatorio Astronomico di Brera, via E.~Bianchi 46, I-23807 Merate, Italy \label{brera}
\and
INAF - Osservatorio di Astrofisica e Scienza dello Spazio, via P.~Gobetti 93/3, I-40129, Bologna, Italy \label{oas}
\and
INFN, Sezione di Bologna, viale Berti Pichat 6/2, I-40127, Bologna, Italy \label{infn}
}

\date{Received XXX; accepted YYY}

\abstract
{Diffuse radio emission associated with the intra-cluster medium (ICM) is observed in a number of merging galaxy clusters. It is currently believed that in mergers a fraction of the kinetic energy is channeled into non-thermal components, such as turbulence, cosmic rays and magnetic fields, that may lead to the formation of giant synchrotron sources in the ICM.}
{Studying merging galaxy clusters in different evolutionary phases is fundamental to understanding the origin of radio emission in the ICM.}
{We observed the nearby galaxy cluster pair RXC J1825.3+3026 ($z\sim0.065$) and CIZA J1824.1+3029 ($z\sim0.071$) at $120-168$ MHz with the \lofarE\ (\lofar) and made use of a deep (240~ks) \xmm\ dataset to study the non-thermal and thermal properties of the system. RXC J1825.3+3026 is in a complex dynamical state, with a primary on-going merger in the E-W direction and a secondary later stage merger with a group of galaxies in the SW, while CIZA J1824.1+3029 is dynamically relaxed. These two clusters are in a pre-merger phase.}
{We report the discovery of a Mpc-scale radio halo with a low surface brightness extension in RXC J1825.3+3026 that follows the X-ray emission from the cluster center to the remnant of a galaxy group in the SW. This is among the least massive systems and the faintest giant radio halo known to date. Contrary to this, no diffuse radio emission is observed in CIZA J1824.1+3029 nor in the region between the pre-merger cluster pair. The power spectra of the X-ray surface brightness fluctuations of RXC J1825.3+3026 and CIZA J1824.1+3029 are in agreement with the findings for clusters exhibiting a radio halo and the ones where no radio emission has been detected, respectively.}
{We provide quantitative support to the idea that cluster mergers play a crucial role in the generation of non-thermal components in the ICM.}


\maketitle
%

\section{Introduction}

\begin{figure*}
 \centering
 \includegraphics[width=.33\textwidth,trim={0.8cm 0cm 0.8cm 0cm},clip]{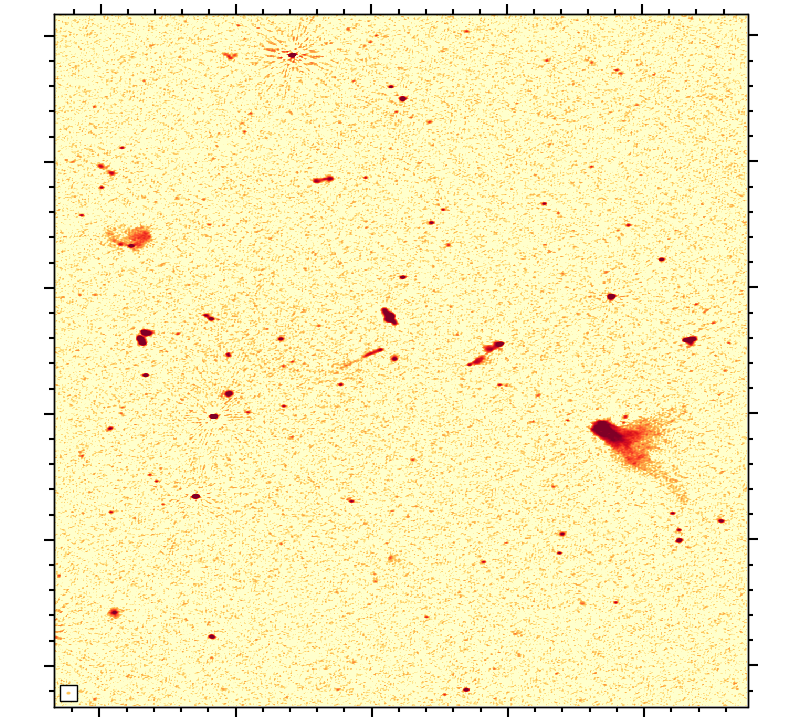}
 \includegraphics[width=.33\textwidth,trim={0.8cm 0cm 0.8cm 0cm},clip]{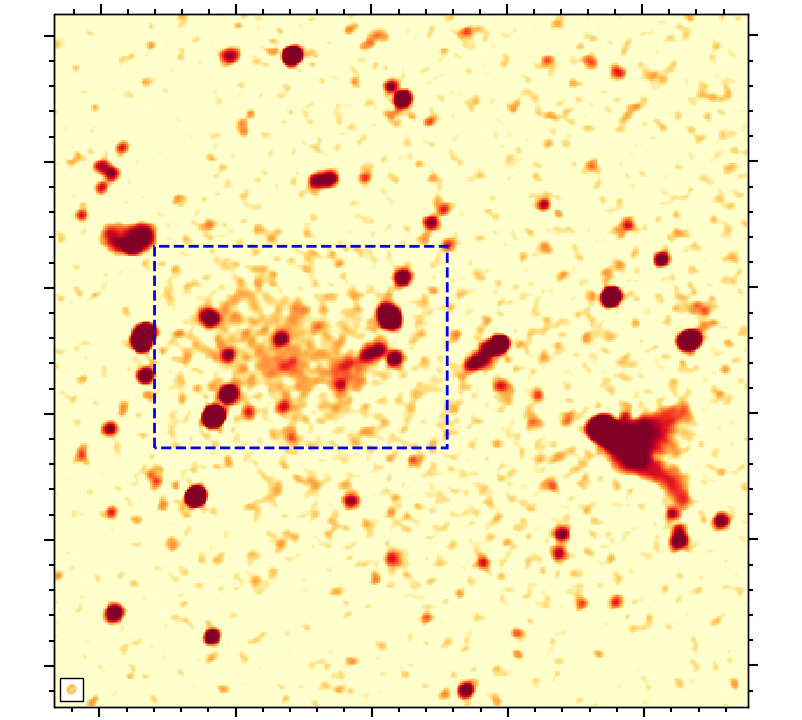}
 \includegraphics[width=.33\textwidth,trim={0.8cm 0cm 0.8cm 0cm},clip]{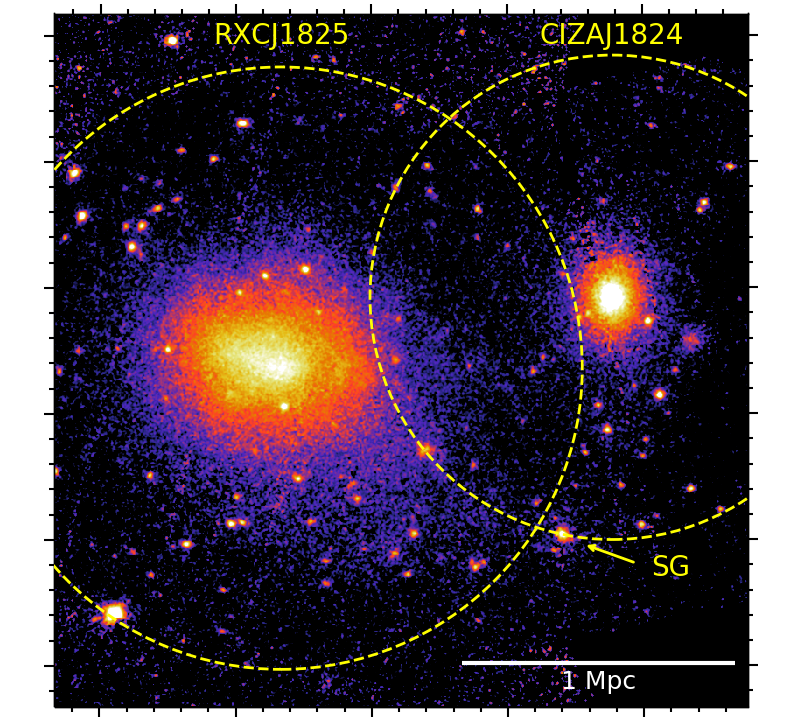}
 \caption{The cluster pair RXCJ1825/CIZAJ1824 as observed with \lofar\ HBA at high (\textit{left}) and medium (\textit{center}) resolution, and with \xmm\ in the $0.5-2.0$ \kev\ band (\textit{right}). The resolution and rms noise of the \lofar\ images are $8.5\arcsec \times 4.7\arcsec$ and $\sigma = 110$ \mujyb\ (high), and $27.1\arcsec \times 24.4\arcsec$ and $\sigma = 220$ \mujyb\ (medium). The beam sizes are shown in the bottom left corners. The blue box in the \lofar\ medium resolution image shows the region where we evaluate the flux density of the halo. Yellow circles in the \xmm\ image denote the approximate location of \rfive\ for each cluster (\cf\ Tab.~\ref{tab:properties}) while the arrow indicates the Southern Galaxy (SG). The displayed images have matched coordinates and cover a FoV of $33\arcmin \times 33\arcmin$ (\cf\ Fig.~\ref{fig:overlay}).}
 \label{fig:view}
\end{figure*}

Galaxy clusters form through the continuous accretion of matter over cosmic time. During accretion, turbulent flows and shock waves are produced in the intra-cluster medium (ICM) and in some circumstances they may generate cluster-scale synchrotron emission that is commonly referred to as giant radio halos or relics \citep[\eg][for a recent review]{vanweeren19rev}. The former category consists of apparently unpolarized sources found at the cluster center with a morphology similar to that of the X-ray emitting gas. The latter are elongated and often highly polarized at decimetric wavelengths and are located in the cluster outskirts. Both halos and relics are generally observed in massive dynamically disturbed clusters \citep[\eg][]{cassano10connection, cassano13, cuciti15} and trace relativistic electrons and magnetic fields distributed in the ICM on Mpc-scales that are eventually (re)accelerated and amplified during merger events \citep[\eg][for a review]{brunetti14rev}. \\
\indent
Observations at low radio frequencies with the \lofarE\ (\lofar) are revealing unprecedented details of the complex diffuse emission from the ICM \citep[\eg][for the most recent works]{birzan19, botteon19a781, clarke19, hoang19a520, hoang19a2146, mandal19, savini19, wilber19}. In particular, \lofar\ is entering into unexplored territories for the study of non-thermal phenomena from the ICM, allowing us to explore the processes undergoing in the very early phases of a merger, well before the core crossing \citep{bonafede18, botteon18a1758, botteon19a781, govoni19}. Theoretical works show that equatorial shocks and turbulent flows should be excited in the ICM between pre-merging clusters; however, the kinetic energy associated with shock and turbulence at this phase is expected to be smaller than in merging systems \citep[\eg][]{vazza17turbulence, ha18shocks} and it is unclear if a significant fraction of this energy can be channeled into non-thermal components already at this stage. \\
\indent
The galaxy clusters RXC J1825.3+3026 and CIZA J1824.1+3029 (hereafter RXCJ1825 and CIZAJ1824, respectively) constitute a binary system (mass ratio 1:1.6) at low redshift also known as Lyra complex \citep{clavico19arx, girardi19arx}. Due to its low Galactic latitude, this system has been challenging to observe but the first focused X-ray \citep{clavico19arx} and optical \citep{girardi19arx} studies have recently been completed (see Tab.~\ref{tab:properties} for the main properties of the two clusters). However, to date, no targeted radio observations have been published. The picture emerging from both X-ray and optical data is that RXCJ1825 and CIZAJ1824 are gravitationally connected but do not have interacted yet, \ie\ they are in a pre-merger phase. Whereas CIZAJ1824 is dynamically relaxed, RXCJ1825 (\ie\ the most massive cluster of the system) shows clear signatures of on-going merging: primarily, two brightest cluster galaxies at the center and an irregular X-ray morphology \citep{clavico19arx}. In addition to this main merger, a minor collision also occurred in the SW periphery of RXCJ1825, where an extension of the X-ray emission is found between RXCJ1825 and the Southern Galaxy (SG in Fig.~\ref{fig:view}, right panel). This is an elliptical galaxy showing evidence for a $\sim 1$ \kev\ ``corona'' surrounded by slightly hotter gas ($\sim 2$ \kev), suggesting that this is the remnant of a group of galaxies that was ram pressure stripped by the interaction with RXCJ1825 and caused the X-ray surface brightness extension in the SW \citep{clavico19arx}. All these features point towards a complex system involving multiple components, making this target an excellent candidate to search for diffuse synchrotron sources in the ICM at different dynamical stages. In particular, we are observing i) a relaxed cluster (CIZAJ1824), ii) an on-going merger (RXCJ1825, in the E-W), iii) a pre-merger (the pair RXCJ1825/CIZAJ1824) and iv) a post-merger (RXCJ1825 and a group of galaxies, in the SW). \\
\indent
In this paper, we report on the results from deep radio observations targeting the Lyra complex using \lofar\ High Band Antennas (HBA). We assume a \lcdm\ cosmology with $\omegal = 0.7$, $\omegam = 0.3$, $\hzero = 70$ \kmsmpc, and adopt the convention $S_\nu \propto \nu^{-\alpha}$ for radio synchrotron spectrum afterwards.

\begin{table}
 \centering
 \caption{Properties of RXCJ1825 and CIZAJ1824. Redshifts are taken from \citet{girardi19arx} while \mfive\ and \rfive\ are derived from the deprojected mass profiles and are taken from \citet[][RXCJ1825]{ettori19} and \citet[][CIZAJ1824]{clavico19arx}. The luminosity within \rfive\ in the $0.1-2.4$ keV band \lfive\ and the average temperature $\langle kT \rangle$ is from \citet{clavico19arx}.}
 \label{tab:properties}
  \begin{tabular}{lrr} 
  \hline
  & RXCJ1825 & CIZAJ1824 \\
  \hline
  Redshift & 0.065 & 0.071 \\
  Right ascension (h, m, s) & 18 25 20.0 & 18 24 06.8 \\
  Declination (\deg, \arcmin, \arcsec) & $+$30 26 11.2 & $+$30 29 32.5 \\
  \mfive\ ($10^{14}$ \msun) & $4.08\pm0.13$ & $2.46\pm0.63$ \\
  \rfive\ (kpc) & $1105\pm12$ & $932\pm79$ \\
  \lfive\ ($10^{44}$ \ergs) & $2.42\pm0.02$ & $0.74\pm0.02$ \\
  $\langle kT \rangle$ (keV) & $4.86\pm0.05$ & $2.14\pm0.05$ \\
  Scale (kpc arcsec$^{-1}$) & 1.248 & 1.354 \\
  \hline
  \end{tabular}
\end{table}

\section{Observations and data reduction}

\subsection{\lofar}

The cluster pair RXCJ1825/CIZAJ1824 was observed with \lofar\ in \texttt{HBA\_DUAL\_INNER} mode on 21 June 2018 (project code: LC10\_013). The observation was performed following the scheme of the \lotssE\ \citep[\lotss;][]{shimwell17}, \ie\ 8~hr on-source time bookended by two 10~min scans of the flux calibrator 3C295 using \lofar\ HBA operating in the $120-168$ MHz frequency band. The 48 MHz bandwidth is centered at the central frequency 144 MHz, which is also the reference frequency for the \lofar\ images shown here. \\
\indent
We make use of the direction-dependent data reduction pipeline\footnote{\url{https://github.com/mhardcastle/ddf-pipeline}} v2.2 developed by the \lofar\ Surveys Key Science Project. This latest version of the pipeline includes improvements in the calibration and imaging of extended sources and is currently adopted to process \lotss\ observations \citep[see Section 5 in][]{shimwell19}, and will be thoroughly discussed in Tasse et al. (in prep.). The data processing exploits \prefactor\ \citep{vanweeren16calibration, williams16, degasperin19}, \killms\ \citep{tasse14arx, tasse14, smirnov15} and \ddfacet\ \citep{tasse18} to perform direction-independent and direction-dependent calibration and imaging of the entire \lofar\ field-of-view (FoV). To improve the image quality towards the target field, we subtract out from the \uv-data all the sources in the \lofar\ FoV using the models derived from the pipeline except those in a $38.3\arcmin \times 38.3\arcmin$ region containing RXCJ1825 and CIZAJ1824. Then, we perform additional phase and amplitude self-calibration loops to correct the residual artifacts in the small extracted region where the direction-dependent errors are assumed to be negligible. The details of this extraction and re-calibration step will be discussed in a forthcoming paper (van Weeren et al., in prep.). \\
\indent
Images at different resolutions are produced with \wsclean\ v2.6 \citep{offringa14} by using the Briggs weighting scheme \citep{briggs95} with \texttt{robust=-0.5} and suitable tapering of the visibilities. An inner \uv-cut of $80\lambda$, corresponding to an angular scale of $43\arcmin$, has been applied to the data to drop the shortest spacings where calibration is more challenging. The \lofar\ high- and medium-resolution images of the clusters are shown in Fig.~\ref{fig:view}. \\
\indent
Due to inaccuracies in the \lofar\ HBA beam model, the \lofar\ flux density scale can show systematic offsets  \citep[see][]{vanweeren16calibration, hardcastle16}. We check and correct the flux density scale by comparing the brightest compact sources extracted from the \tgssE\ \citep[\tgss;][]{intema17} with the \lofar\ image. Throughout the paper, we have applied the correction factor computed from the mean \lofar/\tgss\ integrated flux density ratio of 1.15; uncertainties on the \lofar\ flux densities are dominated by the calibration error of 20\%, which has been adopted in agreement with \lotss\ measurements \citep{shimwell19}.

\subsection{XMM-Newton}

RXCJ1825 was observed in the context of the \xcopE\ \citep[\xcop;][]{eckert19, ettori19, ghirardini19universal} for a total exposure time of 240~ks divided into two central pointings and four offset pointings, one of whose containing CIZAJ1824. We retrieved and processed these observations following standard data reduction recipes of the \xmm\ \esasE\ \citep[\esas;][]{snowden08}. In Fig.~\ref{fig:view} (right panel) we show the \xmm\ background-subtracted and exposure-corrected mosaic image in the $0.5-2.0$ keV band of the cluster pair. For more details about the X-ray data analysis and interpretation, the reader is referred to \citet{clavico19arx}.

\section{Results}

\subsection{Diffuse emission in the ICM}

\begin{figure*}
 \centering
 \includegraphics[width=.49\hsize,trim={0cm 0cm 0.8cm 0cm},clip]{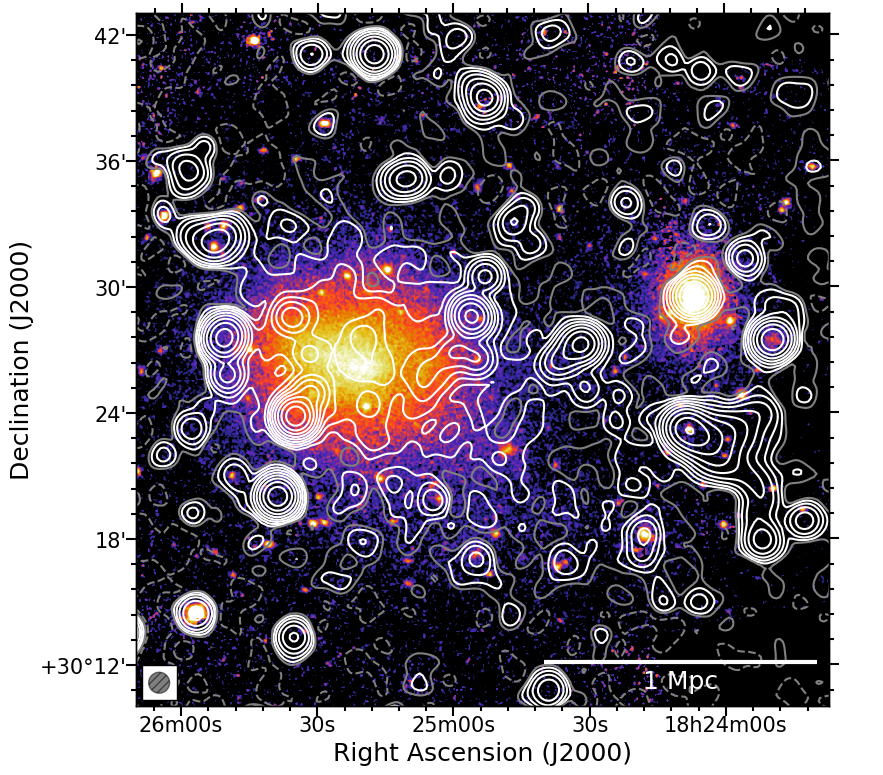}
 \includegraphics[width=.49\hsize,trim={0cm 0cm 0.8cm 0cm},clip]{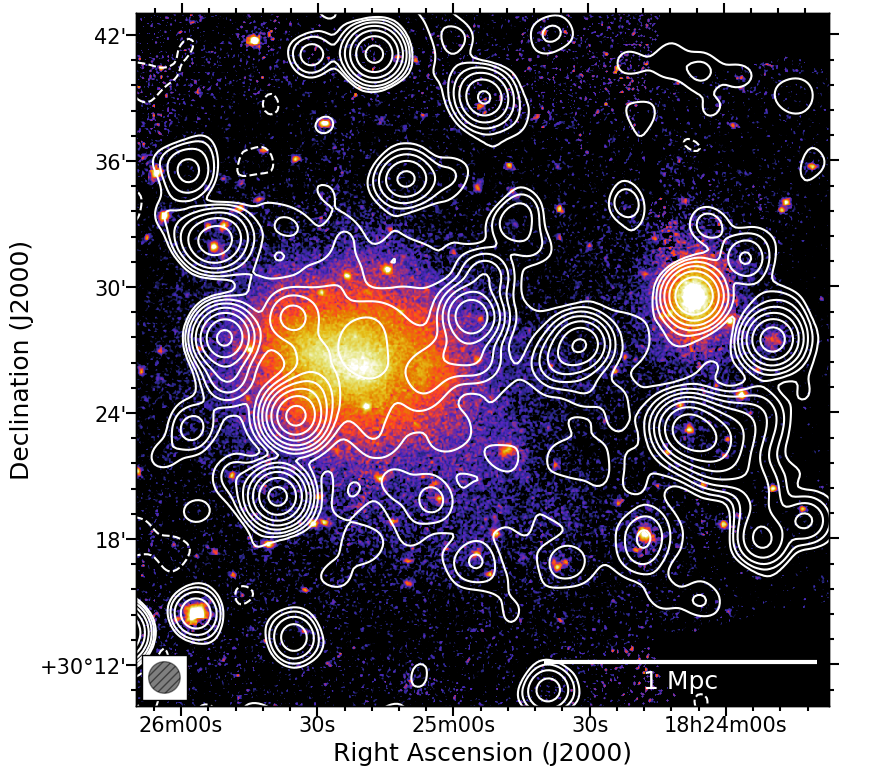}
 \caption{\lofar\ radio contours overlaid on the \xmm\ color image. \textit{Left}: low-resolution ($60\arcsec \times 60\arcsec$) contours spaced by a factor of 2 starting from $1.5\sigma$ (the first contour is reported in gray), where $\sigma = 300$ \mujyb. The negative $-1.5\sigma$ contours are shown in dashed gray. \textit{Right}: very low-resolution ($90\arcsec \times 90\arcsec$) contours spaced by a factor of 2 starting from $3\sigma$, where $\sigma = 415$ \mujyb. The negative $-3\sigma$ contours are shown in dashed. The beams are displayed in the bottom left corners.}
 \label{fig:overlay}
\end{figure*}

The \lofar\ high-resolution image of the Lyra complex shows numerous radio galaxies with a wide range of morphologies (Fig.~\ref{fig:view}, left panel). The most prominent is the tailed radio galaxy that is located between CIZAJ1824 and the Southern Galaxy. This is discussed in Section~\ref{sec:comet}. At a resolution of $27.1\arcsec \times 24.4\arcsec$ (Fig.~\ref{fig:view}, central panel), diffuse radio emission from the ICM of RXCJ1825 is clearly observed. The source follows the X-ray thermal emission and is slightly elongated in the E-W direction. Due to its location in the system, morphology, and cluster-size extent, we classify this source as a giant radio halo. In contrast, most of the radio emission from the vicinity of CIZAJ1824 is associated with the brightest cluster galaxy at its center, as commonly observed in cool-core systems. At this medium resolution, no significant radio emission is detected between RXCJ1825, CIZAJ1824, and the Southern Galaxy. \\
\indent
To better recover the emission from the radio halo in RXCJ1825 and study its connection with the thermal gas, we produce a \lofar\ low-resolution image by applying a Gaussian \uv-taper of 40\arcsec\ and then convolving it with a Gaussian beam of $60\arcsec\times60\arcsec$ ($\sim75~\rm{kpc} \times 75~\rm{kpc}$) in the image plane. In Fig.~\ref{fig:overlay} (left panel) we show the \lofar\ low-resolution contours overlaid on the \xmm\ image. This image shows more clearly the radio halo emission at the cluster center and also suggests the presence of a low surface brightness extension from the halo towards the Southern Galaxy. The halo has a projected size of 1.0 $\times$ 0.8 Mpc, if we take into account the radio emission above the $3\sigma$ level. The halo extension towards the Southern Galaxy is patchy at this confidence level; the $1.5\sigma$ contours highlight better the presence of the underlying low surface brightness extension of emission which remarkably follows the X-ray morphology of the cluster (Fig.~\ref{fig:overlay}, left panel). To improve the signal-to-noise of this feature, we produce a lower resolution image by using a Gaussian beam of $90\arcsec\times90\arcsec$ ($\sim112~\rm{kpc} \times 112~\rm{kpc}$). Fig.~\ref{fig:overlay} (right panel) shows the \lofar\ very low-resolution contours where the extension of the radio halo towards the Southern Galaxy in the SW is detected at $3\sigma$. The linear extent of the diffuse radio emission in this direction is up to $\sim1.8$ Mpc. In Section~\ref{sec:discussion} we discuss the possible origin of this extension, which is likely related to merger between RXCJ1825 and a galaxy group. Finally, we note that in the \lofar\ low-resolution images of Fig.~\ref{fig:overlay}, the radio emission located between RXCJ1825 and CIZAJ1824 is related to a double radio galaxy (\cf\ Fig.~\ref{fig:view}, left panel). \\
\indent
We evaluate the flux density of the radio halo in the blue box in Fig.~\ref{fig:view} (central panel), which roughly covers the $3\sigma$ contour of Fig.~\ref{fig:overlay} (left panel), adopting the two following procedures. \\
As a first approach, we subtract the clean component models of the discrete sources observed at high resolution in Fig.~\ref{fig:view} (left panel) from the visibilities, re-image the new dataset, and evaluate the flux density in the source-subtracted images. In this case, the halo flux density is $S_{144} = 153 \pm 31$ mJy at 144 MHz and is consistent at medium and low resolution. \\
As a second approach, we measure the flux density in the blue box of Fig.~\ref{fig:view} (central panel) taking into account both discrete sources and diffuse emission and then subtract the flux densities of the 18 discrete sources measured in the high-resolution image. This step is useful to evaluate the accuracy of the subtraction of radio galaxies with extended structures (3 out 18, accounting for $\sim208$ m\jy), such as those observed in the direction of RXCJ1825. The total (discrete sources+diffuse) flux density is $\sim885$ m\jy, consistently derived in the medium- and low-resolution images. By subtracting the flux density of $\sim712$ m\jy\ due to the 18 embedded sources, a flux density of $S_{144} = 173 \pm 35$ mJy is associated with the ICM emission. This value is higher than that found in the first approach and highlights the complexity of subtracting the extended emission of the discrete sources embedded in the radio halo. \\
\indent
For the reminder of this paper, we adopt the average flux density value of $S_{144} =163\pm47$ m\jy\ for the radio halo in RXCJ1825. The radio power at 144 MHz is $P_{144} = (1.7\pm0.5) \times 10^{24}$ \whz. If we assume a typical halo spectral index of $\alpha = 1.3$ \citep[\eg][and references therein]{feretti12rev}, the expected flux density at 1.4 GHz is $S_{1.4} = 8.5\pm2.5$ mJy, implying a radio power of $P_{1.4} = (8.7\pm2.5) \times 10^{22}$ \whz. This radio power is a factor of $2-4$ below the extrapolation at the cluster mass (or luminosity) of the best-fit $P_{1.4}-\mfive$ (or $P_{1.4}-\lfive$) relation of \citet{cassano13}. The flux density values expected at 1.4 GHz are below the sensitivity level of our reprocessed \nvssE\ \citep[\nvss;][]{condon98} images, which have an rms of 0.22 \mjyb. However, very short snapshots imply sparse \uv-coverage which is known to prevent the detection of large-scale low surface brightness diffuse emission. Therefore we could not provide a meaningful constraint on the spectral index of the radio halo. Deep, pointed \jvla\ and/or u\gmrt\ observations are planned to perform this kind of analysis. \\
\indent
To the best of our knowledge, RXCJ1825 is the least powerful giant radio halo discovered so far and one of the smallest cluster known to date to host this kind of diffuse emission. It probes regions in the $P_{1.4}-\mfive$ and $P_{1.4}-\lfive$ planes which are poorly constrained due to the limited sensitivity of previous instruments \citep[\eg][]{cassano13, birzan19}. \lofar\ and future radio interferometers (such as the \skaE) have the sensitivity to explore the lower-end of the correlations and to test the theoretical models of radio halo formation in these regimes \citep[\eg][]{cassano10lofar, cassano12}. \\
\indent
We can estimate an upper limit to the flux density of the non-detected diffuse emission in CIZAJ1824 via $S = A\times \sigma$ \citep[\eg][]{hoang18}. If we consider a conservatively large radio halo area of $A=250^2 \pi$ kpc$^2$ and the noise of our low-resolution \lofar\ image, we obtain an upper limit of $S_{144} < 8.1$ m\jy\ for the level of diffuse emission at the center of CIZAJ1824. Assuming, again, a spectral index index of $\alpha = 1.3$, this would imply a radio power more than a factor of 10 below the extrapolation of the $P_{1.4}-\mfive$ relation, underlying the dichotomy between RXCJ1825 (merger, with radio halo) and CIZAJ1824 (relaxed, without diffuse radio emission).

\begin{figure}[ht]
 \centering
 \includegraphics[width=\hsize,trim={0cm 0cm 0cm 0cm},clip]{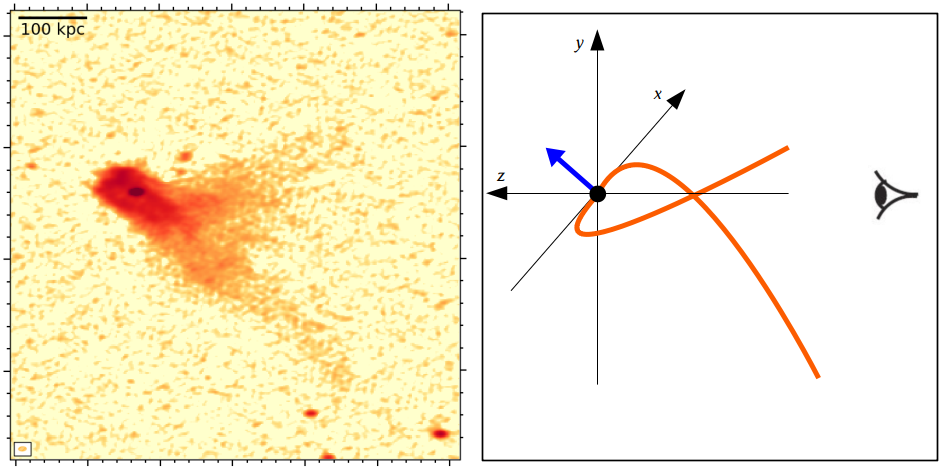}
 \caption{Zoom-in of the tailed radio galaxy in Fig.~\ref{fig:view} (left panel). A cartoon showing the projection that could explain the source morphology is also reported. The galaxy is moving in the direction indicated by the blue arrow.}
 \label{fig:cartoon}
\end{figure}

\subsection{Tailed radio galaxy}\label{sec:comet}

A bright tailed radio galaxy is observed between CIZAJ1824 and the Southern Galaxy. \citet{girardi19arx} found that the host galaxy, which is cospatially coincident with the nucleus of the radio galaxy, has redshift $z=0.0699\pm0.0003$, hence it is part of the Lyra complex, which mean redshift is $\langle z \rangle = 0.0674\pm0.0003$ (based on 198 spectroscopically confirmed members of the system). The flux density of the source is $S_{144} = 577 \pm 115$ mJy. \\
\indent
Tailed radio galaxies are commonly divided into narrow-angle and wide-angle sources, depending on the angle between the radio jets/lobes as they deploy in the ICM \citep[\eg][]{miley80rev}. In the case (Fig.~\ref{fig:cartoon}), the nucleus of the radio galaxy is embedded in a cocoon of emission with a forked diffuse elongation. Due to the lack of high-resolution data, at the moment it is premature to classify this source in one of the two above-mentioned classes. Its morphology could be due to projection effects related to the motion of the host galaxy moving at large angle with respect to the plane of the sky \citep{girardi19arx}, similarly to the case of NGC 7385 \citep{rawes15}. A cartoon indicating the proposed scenario is shown in Fig.~\ref{fig:cartoon}. Distorted radio galaxies can be generated by the interaction between the tails and the surrounding ICM, providing another demonstration of the complexity of the system.

\section{Discussion}\label{sec:discussion}

\begin{figure}
 \centering
 \includegraphics[width=\hsize,trim={0cm 0cm 0cm 0cm},clip]{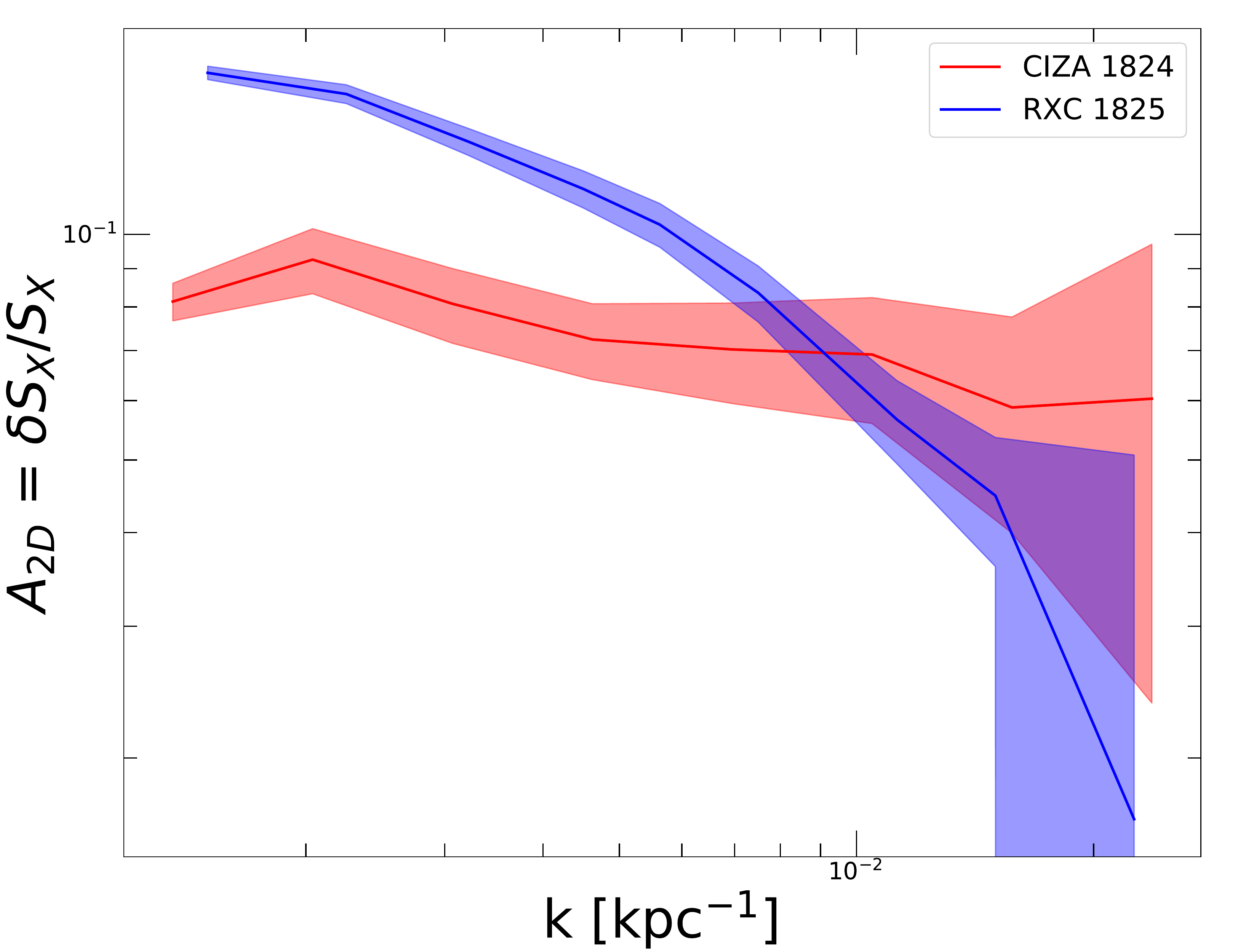}
 \caption{Fractional amplitude of projected (2D) X-ray surface brightness fluctuations $A_{2D}=(P_{2D}2\pi k^2)^{1/2}$ for RXCJ1825 (blue) and CIZAJ1824 (red) as a function of wave number $k$. The power spectra $P_{2D}$ were extracted within a circle of 200 kpc radius around the X-ray peak of both structures. The shaded areas show the $1\sigma$ confidence intervals for both regions.}
 \label{fig:power_spectra}
\end{figure}

\begin{figure*}
 \centering
 \includegraphics[width=.8\hsize,trim={0cm 0cm 0cm 0cm},clip]{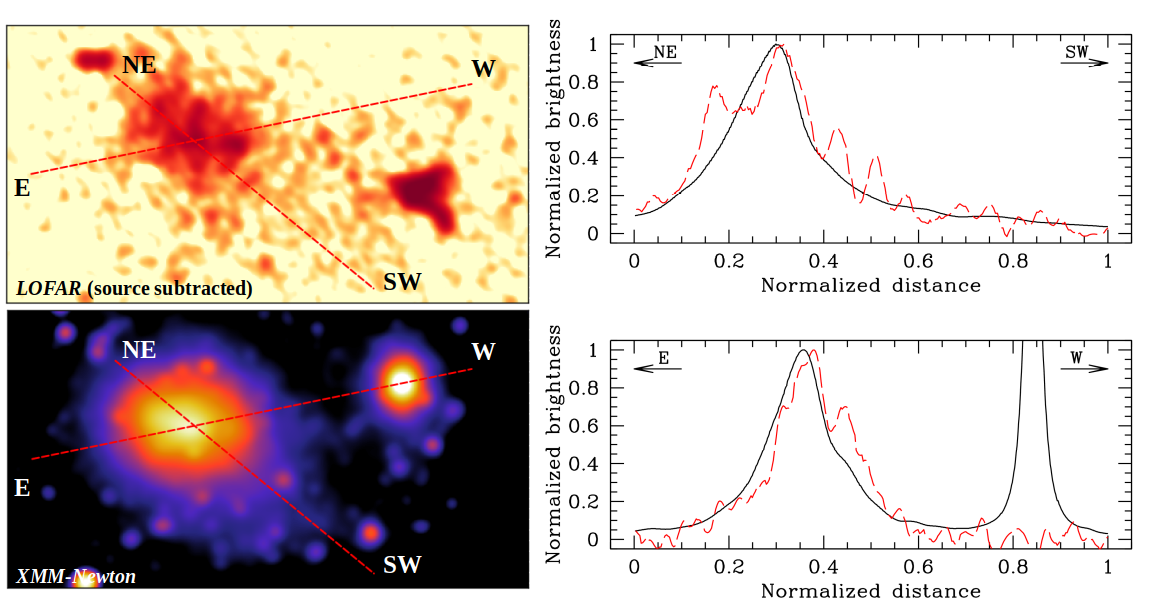}
 \caption{One-dimensional brightness profiles of the X-ray (black lines) and radio emission (red dashed lines) extracted in the dashed lines reported in the left panels whose display the \lofar\ (discrete source subtracted) and \xmm\ images convolved with a comparable resolution of 60\arcsec. The profiles are normalized at the brightness peak of RXCJ1825 in each band.}
 \label{fig:slices}
\end{figure*}

The detection of a radio halo in the dynamically disturbed cluster RXCJ1825 and the absence of diffuse radio emission in the cool-core system CIZAJ1824 strongly support the idea that merger events play a fundamental role in the generation of non-thermal components in the ICM \citep[\eg][]{cassano10connection}. \\
\indent
The Fourier power spectrum analysis of the X-ray surface brightness fluctuations provides information on the Mach numbers of turbulent motions in the ICM \citep[\eg][]{gaspari13, gaspari14, zhuravleva14relation}. We use the deep \xmm\ observation to extract the power spectra of RXCJ1825 and CIZAJ1824 within a radius of 200 kpc from the X-ray peak following the method of \citet{eckert17turbulent}. As shown in Fig.~\ref{fig:power_spectra}, the amplitude of gas density fluctuations $\delta \rho / \rho$ at the maximum scale of the main cluster RXCJ1825 is a factor of $\sim2$ larger than that of the cool-core CIZAJ1824. Since the sound speed $c_s$ in RXCJ1825 is a factor $\sim 1.5$ higher, this would imply that the 3D turbulent velocity dispersion $\sigma_v \approx 3.7 c_s \delta \rho / \rho$ of the main system is $\sim3$ times higher than that of CIZAJ1824 \citep[under the hypothesis of isotropic turbulent motions, see][]{gaspari13}. The values of the 2D amplitude of RXCJ1825 and CIZAJ1824 at the largest scale are in line with those observed for radio halo and non-radio halo clusters, respectively, in the sample analyzed by \citet{eckert17turbulent}. \\
\indent
The indication of the low surface brightness radio extension following the X-ray extension towards the SW direction suggests past interaction between RXCJ1825 and the galaxy group which hosted the Southern Galaxy. This region and that between RXCJ1825 and CIZAJ1824 have been investigated in detail by \citet{clavico19arx} by using the deep \xmm\ observation\footnote{In particular, they (i) modeled the surface brightness distribution of the two clusters assuming two elliptical single-$\beta$ model profiles and produced a residual image by comparing the difference pixel-by-pixel between data and model and (ii) created unsharp masked images with different Gaussian smoothing to search for surface brightness features in the ICM.}. Whereas no residual emission is observed in the region between RXCJ1825 and CIZAJ1824, the region between RXCJ1825 and the Southern Galaxy appears dynamically active. We extract one-dimensional brightness profiles along the slices reported in Fig.~\ref{fig:slices} (left panels) using radio (discrete source subtracted) and X-ray images with a comparable resolution of 60\arcsec\ \citep[\eg][]{shimwell16} to quantitatively show that there is a connection between the thermal and non-thermal emission of the ICM. The locations of the two slices are chosen to avoid contaminating sources in the X-ray image. Although the radio image is naturally more sensitive to local brightness variations (\eg\ due to residual emission left over by subtracted discrete sources), the profiles of Fig.~\ref{fig:slices} (right panels) show an overall good match between the X-ray and radio brightness distribution (apart the absence of radio emission from CIZAJ1824), indicating a tight connection between the two components and possibly a common origin. The X-ray extension towards the SW could be due to gas from the group of galaxies that hosted the Southern Galaxy in the past and that has been stripped by ram pressure during the interaction with RXCJ1825 or to gas belonging to RXCJ1825 generated by tidal interaction with the group. In the first case, the radio emission in this region may be explained as synchrotron emission by particles re-accelerated by the turbulence produced in the wake of the group during its motion towards the cluster outskirts. In the second case, the radio emission could result from the advection of the relativistic and thermal plasma from the cluster center to the SW due to the high-velocity motion of the galaxy group. The X-ray spectral analysis does not allow to discriminate between these two possibilities as the temperature of the ICM surrounding the  ``corona'' of the Southern Galaxy is consistent either with that once belonged to a group, or with that of RXCJ1825 at this radial distance \citep[see][]{clavico19arx}. \\
\indent
\citet{girardi19arx} studied the merger kinematics between RXCJ1825 and CIZAJ1824 adopting the two-body model described in detail by \citet{beers82} and \citet{gregory84} and concluded that the clusters are gravitationally bounded and very likely in an incoming orbit. The fact that no surface brightness nor temperature enhancement is observed in the X-rays between RXCJ1825 and CIZAJ1824 (Fig.~\ref{fig:slices}) suggests that the gas between the clusters has not been compressed and heated yet, indicating that these objects are in a pre-merger phase. We do not find any evidence of diffuse radio emission between RXCJ1825 and CIZAJ1824 in our \lofar\ data, confirming the picture drawn by optical and X-ray analyses. So far, possible diffuse radio emission in the ICM between two galaxy clusters has been reported only in the pairs A1758N-A1758S \citep{botteon18a1758} and A399-A401 \citep{govoni19}.

\section{Conclusions}

We have presented results from a \lofar\ HBA observation of the low redshift galaxy cluster pair RXCJ1825/CIZAJ1824. Recent optical and X-rays studies showed that this system is in a pre-merger phase, CIZAJ1824 is a cool-core cluster, while RXCJ1825 is dynamically disturbed and is undergoing a main merger in the E-W direction and experienced an additional collision with a group of galaxies in the SW. Our findings can be summarized as follows. 

\begin{enumerate}
 \item We discover a giant radio halo in RXCJ1825. The halo has a projected size of 1.0 Mpc $\times$ 0.8 Mpc and an integrated flux density in the range $S_{144} = 163\pm47$ mJy, corresponding to a radio power of $P_{144} = (1.7\pm0.5) \times 10^{24}$ \whz. 
 \item RXCJ1825 is the least powerful radio halo know to date and one of the least massive systems hosting such an object. Assuming a spectral index $\alpha=1.3$, it would fall a factor of $2-4$ below the extrapolation of the current $P_{1.4}-\mfive$ relation in the low-mass regime.
 \item The radio halo has a low surface brightness extension in the direction of the Southern Galaxy, leading to a maximum linear extent of the diffuse radio emission up to $\sim1.8$ Mpc. The remarkable spatial coincidence between the thermal and non-thermal emission indicates that this feature is a consequence of the energy dissipated on small scales due to the interaction between RXCJ1825 and a galaxy group.
 \item The radio emission from CIZAJ1824 comes from the brightest cluster galaxy, as commonly observed in relaxed systems. The highly sensitive \lofar\ observation has allowed us to place an upper limit for the diffuse radio emission a factor of 10 below the extrapolation of the $P_{1.4}-\mfive$ relation.
 \item No diffuse emission is detected between RXCJ1825 and CIZAJ1824, which are in a pre-merger phase.
\end{enumerate}

Overall, the dichotomy observed between the dynamical states, radio properties, and power spectra of X-ray surface brightness fluctuations of the two galaxy clusters paints a consistent picture connecting cluster mergers and the generation of diffuse radio emission in the ICM. \\
\indent
Our results show that \lofar\ has the potential to detect new extended radio sources in the ICM in nearby and low-mass clusters that were previously missed by old generation instruments/surveys. This allows us to open a new window into the study of diffuse emission in these systems that is crucial to constrain the low-power/low-mass end of the $P_{1.4}-\mfive$ scaling relation. 

\begin{acknowledgements}
We thank M. Girardi and W. Boschin for sharing with us details of their optical analysis and the anonymous referee for useful suggestions. RJvW acknowledges support from the VIDI research programme with project number 639.042.729, which is financed by the Netherlands Organisation for Scientific Research (NWO). ABon acknowledges financial support from the ERC-Stg DRANOEL, no 714245, and from the MIUR grant FARE SMS. SE acknowledges financial contribution from the contracts ASI 2015-046-R.0. SE and MS acknowledge financial contribution from the contract ASI-INAF n.2017-14-H.0. HJAR acknowledge support from the ERC Advanced Investigator programme NewClusters 321271.
This paper is based (in part) on data obtained with the International LOFAR Telescope (ILT) under project code LC10\_013. LOFAR \citep{vanhaarlem13} is the LOw Frequency ARray designed and constructed by ASTRON. It has observing, data processing, and data storage facilities in several countries, which are owned by various parties (each with their own funding sources), and are collectively operated by the ILT foundation under a joint scientific policy. The ILT resources have benefitted from the following recent major funding sources: CNRS-INSU, Observatoire de Paris and Universit\'{e} d'Orl\'{e}ans, France; BMBF, MIWF-NRW, MPG, Germany; Science Foundation Ireland (SFI), Department of Business, Enterprise and Innovation (DBEI), Ireland; NWO, The Netherlands; The Science and Technology Facilities Council, UK; Ministry of Science and Higher Education, Poland; Istituto Nazionale di Astrofisica (INAF), Italy. This research made use of the Dutch national e-infrastructure with support of the SURF Cooperative (e-infra 180169) and the LOFAR e-infra group, and of the LOFAR IT computing infrastructure supported and operated by INAF, and by the Physics Dept. of  Turin University  (under the agreement with Consorzio Interuniversitario per la Fisica Spaziale) at the C3S Supercomputing Centre, Italy. The J\"{u}lich LOFAR Long Term Archive and the German LOFAR network are both coordinated and operated by the J\"{u}lich Supercomputing Centre (JSC), and computing resources on the Supercomputer JUWELS at JSC were provided by the Gauss Centre for Supercomputing e.V. (grant CHTB00) through the John von Neumann Institute for Computing (NIC).  This work is also based on observations obtained with \xmm, an ESA science mission with instruments and contributions directly funded by ESA Member States and NASA. This research made use of APLpy, an open-source plotting package for Python \citep{robitaille12}.
\end{acknowledgements}

%
%

\bibliographystyle{aa}
\bibliography{library.bib}

\end{document}